\documentclass[english,aps,prl,superscriptaddress,floatfix,notitlepage,reprint,show pacs]{revtex4-2}
\usepackage[utf8]{inputenc}
\usepackage[T1]{fontenc}
\usepackage{physics}
\usepackage{natbib}
\usepackage{amsthm}
\usepackage{float}
\usepackage{amssymb}
\usepackage{dsfont}
\usepackage{amsmath}
\usepackage{bm}
\usepackage{sublabel}
\usepackage{latexsym}
\usepackage{sidecap}
\usepackage{placeins}
\usepackage{url}
\makeatletter
\theoremstyle{plain}

\theoremstyle{plain}

\theoremstyle{plain}
\newtheorem*{prop*}{\protect\propositionname}

\usepackage{braket}
\usepackage{txfonts}
\usepackage{pifont}
\usepackage{graphicx}
\usepackage[usenames,dvipsnames]{xcolor}
\usepackage{hyperref}
\usepackage{cleveref}
\hypersetup{
    colorlinks=true,
    linkcolor=Red,       
    citecolor=blue,      
    urlcolor=cyan      
}
\usepackage{orcidlink}
\usepackage{tikz}
\usetikzlibrary{patterns,decorations.text,decorations.pathreplacing,decorations.pathmorphing}
\usepackage{caption}
\usepackage{subcaption}
\usepackage{rotating}
\usepackage{lipsum}

\date{\today}
\newcommand{\beq}{\begin{equation}}
\newcommand{\eeq}{\end{equation}}
\newcommand{\beqa}{\begin{eqnarray}}
\newcommand{\eeqa}{\end{eqnarray}}

\newcommand{\red}{\protect\tikz[baseline=-0.5ex]\draw[red] (0,0)--(0.5,0);}
\newcommand{\blue}{\protect\tikz[baseline=-0.5ex]\draw[blue] (0,0)--(0.5,0);}
\newcommand{\green}{\protect\tikz[baseline=-0.5ex]\draw[green] (0,0)--(0.5,0);}

\newcommand{\blackdashed}{\protect\tikz[baseline=-0.5ex]\draw[black,dashed] (0,0)--(0.5,0);}
\newcommand{\bluedashed}{\protect\tikz[baseline=-0.5ex]\draw[blue,dashed] (0,0)--(0.5,0);}

\begin{document}
\title{Coherence Resonance Phenomena in an Atom-Dimer Two Mode BECs}
\author{Avinaba Mukherjee \orcidlink{0009-0000-3765-6466}}
\thanks{\href{mailto:amphy_rs@caluniv.ac.in}{amphy\_rs@caluniv.ac.in}}
\address{Department of Physics, University of Calcutta, $92$ A. P. C. Road, Kolkata $700009$, India}
\begin{abstract}
We investigate the non-equilibrium dynamics of atomic-molecular Bose-Einstein condensates coupled via a Feshbach resonance, with Gaussian white noise acting on both the coupling strength and the detuning. Using the bosonic Josephson-junction and Bloch-sphere formalisms, we examine how coherence (coupling) noise and imbalance (detuning) noise modify the coherence-resonance point as functions of the initial polarization and the Feshbach detuning. Noise plays a more coherent role when its characteristic timescale matches the intrinsic timescales of the dynamics, leading to extrema in the time-averaged purity and the Husimi-$Q$ distribution. 
 
\end{abstract}
\maketitle
\section{Introduction}
When the characteristic time scale of noise matches that of the external driving force, stochastic resonance (SR) can emerge \cite{witthaut2009dissipation}. In such situations, noise does not overwhelm the dynamics but instead reinforces the effect of the driving field \cite{liu2019symmetry}. This phenomenon improves the signal to noise ratio, thereby enhancing the sensitivity of signal transmission. Equivalently, SR can be viewed as an enhancement of synchronization \cite{han2024semiconductor}. The degree of synchronization is greatest when the signal-to-noise ratio attains its maximum value \cite{fan2017stochastic}. In essence, SR boosts weak signals by converting part of the noise energy into useful signal energy \cite{zhang2022tri}. SR occurs in bistable systems, where noise-induced transitions between two metastable states are assisted by an energy barrier, a weak external periodic signal, and noise \cite{sen2001quantum}.

A fundamental distinction exists between quantum and classical SR: the former is driven by noise of quantum origin, while the latter is induced by thermal fluctuations. If the uncertainty in the initial state is modeled as shot noise, Quantum Stochastic Resonance (QSR) emerges when the system's coherence is maximized at a particular noise strength \cite{witthaut2012stochastic}. Vacuum fluctuations can give rise to QSR at zero temperature \cite{qiu2017optical,xie2018interference}. The SR phenomenon was first observed in the signal processing field in a trigger circuit \cite{fauve1983stochastic}. SR phenomena have been obtained in various types of systems: (i) monostable \cite{agudov2010stochastic}, (ii) bistable \cite{kasai2018divergence}, (iii) multistable \cite{fan2017stochastic}, (iv) periodic potential \cite{jiao2017vibrational}, (v) spin-boson system \cite{grifoni1996coherent}, and (vi) cavity optomechanical systems \cite{monifi2016optomechanically}. 

Coherence resonance (CR) can be regarded as a special case of SR where no external periodic drive is present. CR, in contrast, describes the situation where oscillations become more regular and pronounced in the presence of noise than in its absence \cite{jacobo2010effects}. CR was first introduced by Pikovsky and Kurths \cite{pikovsky1997coherence}. Here, noise plays an organizing role \cite{zhu2022phase}, producing quasi coherent oscillations \cite{yu2018noise} that highlight the coherent influence of quantum fluctuations \cite{kato2021quantum}. CR typically arises in nonlinear systems \cite{fan2023enhancement}. CR plays an important role in opto mechanical systems \cite{yu2018noise}, weak signal detection \cite{aldana2014detection}, stability analysis \cite{agudov2020nonstationary}, spin squeezing systems \cite{sonar2018squeezing}, the Dicke model \cite{witthaut2012stochastic}, and the Jaynes Cummings model \cite{qiu2017optical}. 

In dissipative regimes, the dynamics is described by a Lindblad master equation, while weak system-bath coupling can be modeled as Gaussian white noise \cite{dutta2025introduction}. In double-well Bose-Einstein-Condensates (BECs), noise in hopping and detuning induces damped oscillations of population imbalance and phase coherence toward steady state \cite{stochastic_bosonic_josephson_junction}. Similar ideas apply to atomic–molecular BECs, whose distinct nonlinearity leads to richer bimodal dynamics.\\
In this work, we study a noisy atom--molecule two-level system using its Bloch-sphere representation. We compare two characteristic time scales: the intrinsic time scale of the system and the time scale associated with the noise. When these two time scales become comparable, CR emerges. To identify the CR point, we employ two indicators (Purity, and Husimi-Q distribution function) and determine the noise strength at which they attain their extremal values, thereby confirming the onset of coherence resonance and the regime in which noise plays a coherent role. Finally, we examine the dependence of the intrinsic dynamical timescale, and the resonant coherent noise strength on the initial polarization and the Feshbach detuning, considering noise applied independently to the Feshbach coupling and the detuning. By characterizing these quantities across the relevant parameter space, we determine the conditions under which the interplay between intrinsic dynamics and external fluctuations leads to enhanced coherence, manifested as noise-induced resonant behavior. As our work does not include any external drive, it belongs to the CR domain.
\begin{figure}
\centering
\begin{subfigure}{0.8\linewidth}
    \centering
    \includegraphics[width=\linewidth]{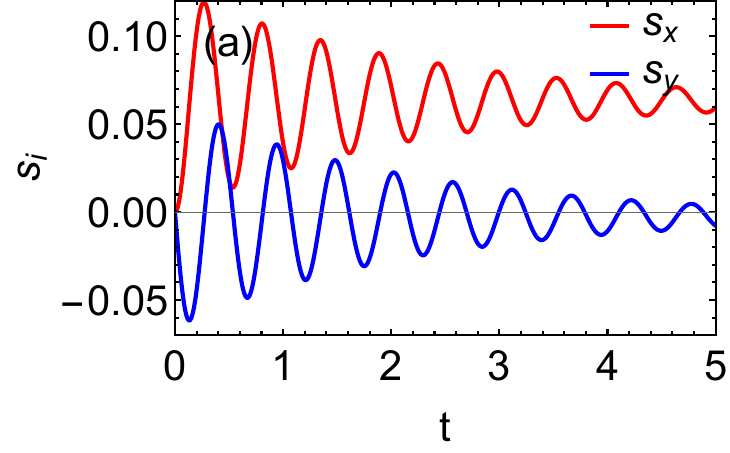}
    \phantomcaption 
    \label{coherence_dynamics}
\end{subfigure}
\begin{subfigure}{0.8\linewidth}
    \centering
    \includegraphics[width=\linewidth]{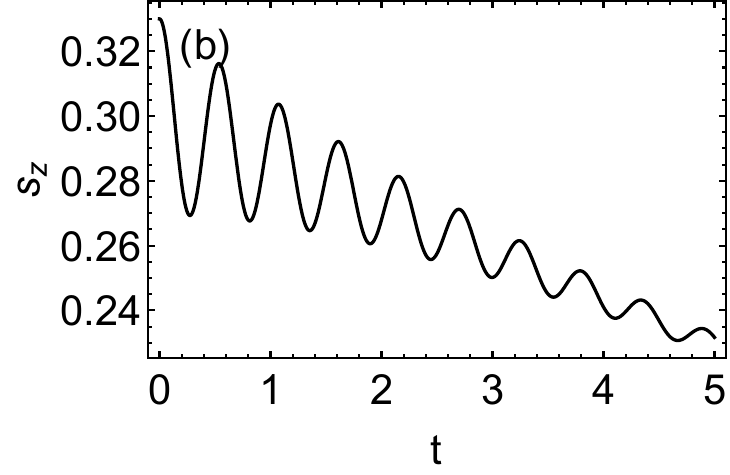}
    \phantomcaption 
    \label{imbalance_dynamics}
\end{subfigure}
\caption[short description]{Relaxation dynamics of the Bloch vector components ($s_i$): (a) $s_x$ (\red) and $s_y$ (\blue), and (b) $s_z$, in the presence of both coupling noise ($\gamma_x$) and detuning noise ($\gamma_z$) acting simultaneously.
}
\label{relaxation s}
\end{figure}
\\
The structure of this paper is as follows. In Sec. \ref{framework}, we introduce the basic model and derive the dynamics of both coherence and polarization within the framework of a two-state Bloch-sphere model. In Sec. \ref{interplay}, we determine the intrinsic dynamical time scale and examine its dependence on the two control parameters of the system: the initial polarization and the Feshbach detuning. In Sec. \ref{section influence}, we determine the range of noise strengths over which noise regularizes the dynamics. In Sec. \ref{5th_section}, we examine how the coherent noise strength varies with the system's two tuning parameters. Finally, we summarize our findings in Sec. \ref{conclusion}.

\section{Basic Hamiltonian Framework and Noise Effects}\label{framework}
A Bloch-vector description of the system is developed in Sec. \ref{toy}, and its behavior in a noisy environment is analyzed in Sec. \ref{TSM}. The relaxation dynamics of the Bloch components are examined in Sec. \ref{relaxation}.
\subsection{Toy Model Bridging Theory and Experiment}\label{toy}
We investigate a system in which two bosonic atoms can coherently bind to form a bosonic molecule via Feshbach resonance \cite{AMBEC,AMBEC2,AMBEC3}. This process is modeled by a two channel framework consisting of an open (entrance) channel and a closed channel \cite{review}. Resonant coupling occurs when the energy levels of these channels coincide, enabling the formation of a bosonic dimer from two free atoms. The energy difference between the atomic BEC (A-BEC) and the molecular BEC (M-BEC) is denoted by $\epsilon_b$ and can be precisely controlled via an external magnetic field.

The coupled atom molecule dynamics are governed by the Hamiltonian \cite{similar_hamiltonian1,similar_hamiltonian2,similar_hamiltonian3,similar_hamiltonian4}
\begin{equation}
\label{hamiltonian}
\begin{split}
\hat{H}= & \frac{u_1}{2V} \hat{a}^\dagger \hat{a}^\dagger \hat{a}\hat{a}+ \frac{u_2}{2V} \hat{b}^\dagger \hat{b}^\dagger \hat{b}\hat{b}+ \frac{u_3}{V} \hat{a}^\dagger \hat{b}^\dagger \hat{b}\hat{a} \\&+ \frac{g}{\sqrt V} (\hat{a}^\dagger \hat{a}^\dagger \hat{b}+ \hat{b}^\dagger \hat{a}\hat{a}) + \epsilon_b \hat{b}^\dagger \hat{b},
\end{split}
\end{equation}
where $\hat{a}^\dagger$ ($\hat{a}$) and $\hat{b}^\dagger$ ($\hat{b}$) are the creation (annihilation) operators for atoms and molecules, respectively. Parameters $u_1$ and $u_2$ describe atom-atom and molecule-molecule interactions, $u_3$ captures atom-molecule interactions, and $g$ is the Feshbach coupling strength. Here, the quantization volume, $V$, is related to the quantization length ($L_0$) through $V \approx L_0^3$. 

Here, both the A-BEC and M-BEC are modeled within a single-mode condensate approximation. The Hamiltonian in Eq. (\ref{hamiltonian}) may be regarded as an effective toy model that retains the key physics of harmonically confined BECs in the weakly interacting regime, while disregarding spatial inhomogeneities arising from the external trapping potential as well as thermal occupation of higher excited modes. Conceptually, this is analogous to the use of a single macroscopic field variable in the Gross–Pitaevskii framework \cite{P-527} to describe trapped condensates, an approach known to reproduce qualitatively reasonable agreement with experimental observations \cite{savage2003bose,albiez2005direct}. A similar level of approximation has also been employed in experimental studies of noisy double-well condensates, where the theoretical description was formulated through a two-mode BEC model \cite{gati2006primary}, closely paralleling the framework adopted in the present work.
\subsection{Bloch Vector Description: Interplay Between Intrinsic Dynamics and Noise}\label{TSM}
To visualize the dynamics, we employ a Bloch vector formalism analogous to spin systems. The two relevant basis states fully molecular and fully atomic map to the north and south poles of the Bloch sphere, respectively. The pseudo-spin (Schwinger) operators \cite{P70,P189} are defined as \cite{bloch4,commutator}:
 $\hat{L}_x=\sqrt{2} (\hat{a}^\dagger \hat{a}^\dagger \hat{b} + \hat{b}^\dagger \hat{a} \hat{a})/N^{3/2}$, $\hat{L}_y =\sqrt{2}i (\hat{a}^\dagger \hat{a}^\dagger \hat{b} - \hat{b}^\dagger \hat{a} \hat{a})/N^{3/2}$, $\hat{L}_z =(2\hat{b}^\dagger \hat{b}-\hat{a}^\dagger \hat{a})/{N}$, and 
$N = 2\hat{b}^\dagger \hat{b}+\hat{a}^\dagger \hat{a}$.    
Here, $\hat{L}_{x}$, and $\hat{L}_{y}$, encode the real and imaginary parts of atom molecule coherence, while $\hat{L}_z$ represents the population imbalance \cite{cui2012atom}. The expectation values of the Bloch vector components are defined as
$s_i = \langle \hat{L}_i \rangle$. 

It is worth emphasizing that the Bloch-vector variables in the present two-mode atom-dimer system do not obey the usual $\mathrm{SU}(2)$ spin algebra encountered in double-well condensates \cite{Linblad_master_equation,stochastic_bosonic_josephson_junction}; rather, they satisfy an $\mathrm{SU}(1,1)$ algebra \cite{khripkov2011quantum}. This distinction originates from the large-$N$ constraint, $s_x^2+s_y^2=(1+s_z)(1-s_z)^2/2$, which confines the dynamics to a generalized Bloch sphere \cite{bloch4}.

Defining scaled parameters
$U_1=Nu_1/V,U_2=Nu_2/V,U_3=Nu_3/V$, and $\tilde{g}=g\sqrt{N/V}$. Including these stochastic perturbations, the large-$N$ form of Eq. (\ref{hamiltonian}) becomes
\begin{equation}
\label{hamiltonian_large_N}
\begin{split}
\hat{\mathcal{H}} = \frac{\hat{H}}{N}
= & \frac{U_1}{8}(\hat{L}_z - 1)^2+
\frac{U_2}{32}(\hat{L}_z + 1)^2 
\\& - \frac{U_3}{8}(\hat{L}_z^2 - 1)+
\frac{\tilde{g} + n_x}{\sqrt{2}} \hat{L}_x+
\frac{\epsilon_b + n_z}{4}(\hat{L}_z + 1),
\end{split}
\end{equation}

with $n_x$ and $n_z$ representing fluctuations in the Feshbach coupling and detuning, respectively. They are modeled as Gaussian white-noise processes, $n_i(t)=\mathrm{d}w_i/\mathrm{d}t$, where $w_i(t)$ are independent Wiener processes satisfying $\langle \mathrm{d}w_i\mathrm{d}w_j\rangle=\gamma_i\delta_{ij}\mathrm{d}t/2$ \cite{stochastic_bosonic_josephson_junction}, implying that $n_i(t)$ are zero-mean, delta-correlated noise sources \cite{noise_property}. Adopting the large-$N$ limit is natural here, since magneto-optical trap experiments typically involve condensates with $10^5$–$10^6$ particles \cite{strecker2003conversion,burt1997coherence}.

 We include only external noise, with non-operator stochastic terms, thereby accounting for fluctuations and decoherence but excluding particle loss. The system thus remains closed with conserved total particle number ($\dot N=0$), an approximation justified by negligible two- and three-body losses in the present system \cite{quantumgas2}.

 In atomic–molecular BECs, such noise commonly arises from condensate interactions with thermal atoms \cite{bloch4,cui2012atom}, where the condensed mode forms the system and the noncondensed fraction acts as a bath, justifying a Gaussian white-noise description \cite{anglin1997cold,ruostekoski1998bose,burt1997coherence}. Thermal scattering induces phase diffusion with rate $\gamma_x$, effectively perturbing the Feshbach coupling. Additional controlled noise may originate from laser intensity and beam-pointing fluctuations, known to cause heating and decoherence \cite{savard1997laser,gardiner2000evaluation}. Fluctuations in detuning ($\gamma_z$), meanwhile, arise from magnetic-field noise near the Feshbach resonance \cite{bloch_vector4}, and thermal fluctuations in the condensate \cite{saha2023phase}.  For ${}^{87}\mathrm{Rb}$, the atomic mode is energetically preferred over the molecular mode \cite{motohashi2010particle}, driving molecular fraction, $2N_b/N\to0$ during evolution.

\subsection{Relaxation Dynamics}\label{relaxation}
Considering only the first moments, the evolution equations in the large-$N$ limit are given by
\begin{subequations}
\label{covariance}
    \begin{equation}
    \label{covariance1}
    \begin{split}
    \dot{s_x}=(2c_1s_z+c_2)s_y-\frac{\gamma_z s_x}{2}
         \end{split}
    \end{equation}
    \begin{equation}
    \label{covariance2}
\begin{split}
    \dot{s_y}=&-(2c_1s_z+c_2)s_x-\frac{\tilde{g}}{\sqrt{2}}(1+2s_z-3s^2_z)\\&+2\gamma_xs_y(3s_z-
1)-\frac{\gamma_z s_y}{2}
        \end{split}
    \end{equation}
    \begin{equation}
    \label{covariance3}
    \begin{split}
    \dot{s_z}=2\sqrt{2}\tilde{g} s_y-\gamma_x(1+2s_z-3s^2_z).
       \end{split}
    \end{equation}
    \end{subequations}
 Here, the effective interaction and detuning are given by $c_1=U_3/2\hbar-U_1/2\hbar-U_2/8\hbar$ and $c_2=U_1/\hbar-U_2/4\hbar-\epsilon_b/\hbar$, both determined by the experimental parameters $u_1$, $u_2$, $u_3$, and $\epsilon_b$.  Notably, when noise is incorporated solely through the coupling and Feshbach detuning, Eq. (\ref{covariance}) takes a form structurally similar to that obtained within the Lindblad formalism in \cite{bloch4}. The real-time dynamics of $s_i$ are shown in Fig. (\ref{relaxation s}).

Eq. (\ref{covariance}) describes a driven damped oscillator in which the driving force arises intrinsically from the system dynamics \cite{quantumgas2}. In the absence of any external periodic driving, the phenomenon belongs to the CR domain.

In Sec. \ref{interplay}, we examine how the intrinsic time scale of the two-level system depends on the initial polarization and the Feshbach detuning.
\section{Intrinsic Time Scale of the Dynamics and Its Variation}\label{interplay}
In Sec. \ref{energy or particle transfer}, we describe the initial conditions for the dynamics and the parameters extracted from available experimental data for ${}^{87}\mathrm{Rb}$. In Sec. \ref{time scale}, we examine the dependence of the intrinsic time scale on the initial polarization and the Feshbach detuning.
\subsection{Initial States and Parameter Regimes}{\label{energy or particle transfer}}
For the atomic-molecular Fock state $\ket{N-2N_b,N_b}$ \cite{khripkov2011quantum}, the Bloch vector components are
\begin{equation}
s_x=s_y=0,
\quad\text{and}\quad
s_z=\frac{4N_b}{N}-1.
\end{equation}
The numerical calculations employ ${}^{87}\mathrm{Rb}$ parameters motivated by atomic–molecular BEC experiments \cite{papp2006observation,zhang2021transition,BJJ1}, with details provided in Table \ref{1st table}. For the parameter set $U_1:U_2:U_3:\tilde g:\epsilon_b = 1:4.3:-7:2:9.1$ (see Table \ref{3rd table}), the corresponding values are $c_1=-4.5$ and $c_2=-9$.

We take Signal to noise ratio, $\mathrm{SNR}\sim10$ for both $\gamma_x$ and $\gamma_z$, consistent with experimental bounds and stochastic resonance studies showing $\sim10\%$ noise strongly affects relaxation while preserving core dynamics \cite{witthaut2008dissipation,wellens2003stochastic}. Comparable fluctuation levels have also been reported in related ultracold-atom experiments \cite{sadgrove2005effect,gati2006noise}. Comparable noise levels are common in resonant ultracold-atom experiments. For example, kicked-rotor studies showed quantum resonance surviving even under large amplitude noise \cite{sadgrove2005effect}, while two-mode BEC interferometry reported phase fluctuations corresponding to roughly $5$–$15\%$ noise \cite{gati2006noise}. These support the experimental relevance of the noise strengths used here (details in Table \ref{4th table}). Details about these parameters value in \hyperref[appendix]{Appendix}.

\subsection{Intrinsic Time Scale of this Dynamics}{\label{time scale}} 
If the relaxation dynamics is described in the state-space representation, then
\begin{subequations}
\begin{equation}
\label{jacobian}
    \dot{\mathbf{s}} =J \mathbf{s},
\end{equation}
where the Jacobian matrix \cite{greiner2003classical} is
\begin{equation}
J=
\begin{pmatrix}
-\frac{\gamma_z}{2} & 2c_1s_z+c_2 & 2c_1s_y\\
-(2c_1s_z+c_2) &
\bigg(2\gamma_x(3s_z-1)-\frac{\gamma_z}{2}\bigg) &
-\bigg(2c_1s_x+\sqrt{2}\tilde{g}(1-3s_z)\bigg)\\
0 & 2\sqrt{2}\tilde{g} &
2\gamma_x(3s_z-1)
\end{pmatrix}.
\end{equation}

\text{Linearizing $J$ about the fixed point}, {we obtain the characteristic equation}
\begin{equation}
    \delta\dot{\mathbf{s}}=J_{\text{linearized}} \delta\mathbf{s}
\end{equation}
\text{the spectrum consists of one real eigenvalue and a pair of} { complex-conjugate eigenvalues. The Liouvillian or spectral gap is defined as the eigenvalue with the least negative real part. The asymptotic relaxation time is then given by the inverse of the absolute magnitude of its real part \cite{fazio2025many}, as shown in}

\begin{equation}
T_i=\frac{1}{|\operatorname{Re}(\lambda_i)|}.    
\end{equation}
 \end{subequations}
 The characteristic times $T_x$ and $T_z$ are obtained when only $\gamma_x$ and only $\gamma_z$ are activated, respectively.
\subsubsection{Initial Polarization Dependence of the Dynamical Timescale}
$T_i$ denotes the recovery time of the system after a perturbation is applied, i.e., the characteristic time over which the perturbation decays and the system relaxes back to the stable equilibrium point. The decay of a perturbation requires the real part of the eigenvalue (the spectral gap), $\lambda_i^R$, to be negative, since the perturbation evolves as $\delta s_i(t)=\delta s_i^0 e^{\lambda_i^R t}$, where $\delta s_i^0$ denotes the initial amplitude of the fluctuation.

\begin{figure}
\centering
\begin{subfigure}{0.9\linewidth}
    \centering
    \includegraphics[width=\linewidth]{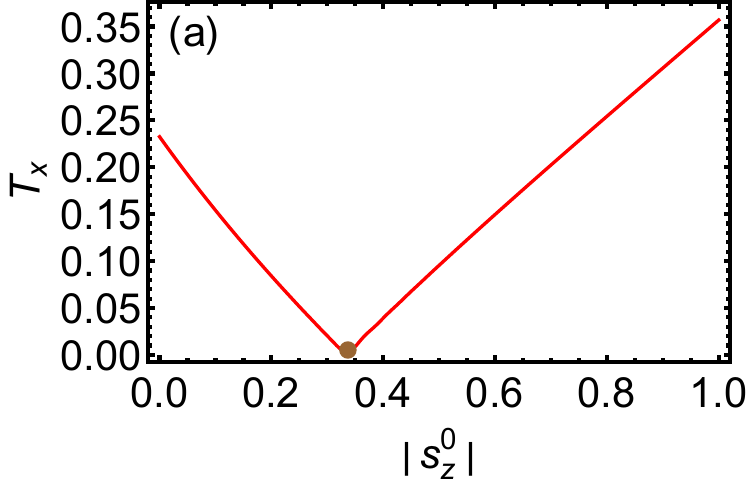}
    \phantomcaption 
    \label{Tx_sz}
\end{subfigure}
\begin{subfigure}{0.9\linewidth}
    \centering
    \includegraphics[width=\linewidth]{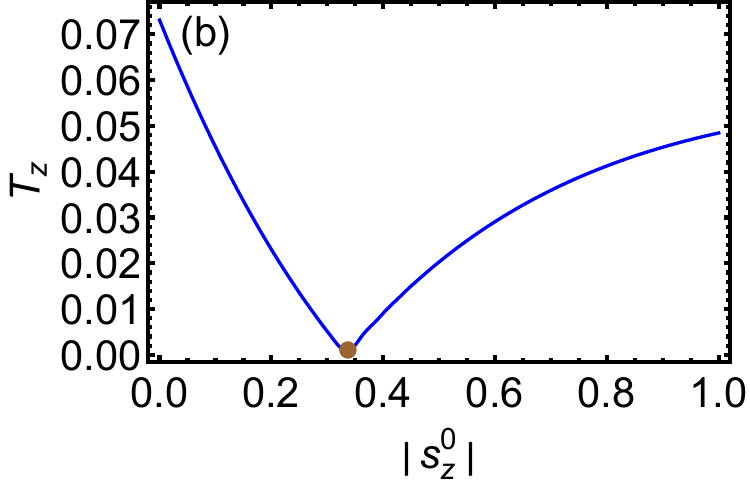}
    \phantomcaption 
    \label{Tz_sz}
    \end{subfigure}
\caption[short description]{Dependence of system time scale, $T_{x(z)}$ on the Initial polarization, ($\lvert s^0_z\rvert$). The (\red) and (\blue) curves correspond to the cases where only (a) coupling noise, $\gamma_x$ and only (b) detuning noise, $\gamma_z$ are activated, respectively. The equilibrium position is highlighted using (\textcolor{brown}{$\bullet$}).}
\label{dynamic time_imbalance}
\end{figure}

Fig. (\ref{dynamic time_imbalance}) shows that the characteristic time $T_i$ increases as the system deviates from an equal population distribution between the two species. The minimum value of $T_i$ occurs near the stable equilibrium point, where the spectral gap attains its maximum. At the fixed point $(0,0,-1/3)$, a finite detuning is present, while the inter-species coupling is strongest when the two species have equal population weights. The combined effect of finite detuning and maximal coupling produces the largest separation between the eigenmodes, resulting in the maximum spectral gap.

From a dynamical perspective, the drift velocity approaches zero at the stable equilibrium point, i.e., $\lvert \dot{s}_z\rvert \rightarrow 0$ \cite{quantumgas2}, indicating that transitions between the two states are strongly suppressed. Equivalently, in terms of mode separation, the equilibrium point corresponds to the maximum energy difference between the two modes. Since the characteristic time scale is inversely related to this energy separation, the largest spectral gap leads to the shortest characteristic time.

In Sec. \ref{noise section}, we explain why $\gamma_x$ is significantly smaller than $\gamma_z$. Therefore, decoherence effects are weaker for $T_x$ in Fig. (\ref{Tx_sz}) than for $T_z$ in Fig. (\ref{Tz_sz}), allowing the relaxation time scale associated with $T_x$ to remain larger over $T_z$.
 
\subsubsection{Feshbach detuning as a Tuning Parameter for Intrinsic Time Scale}
\begin{figure}
\centering
\begin{subfigure}{0.9\linewidth}
    \centering
    \includegraphics[width=\linewidth]{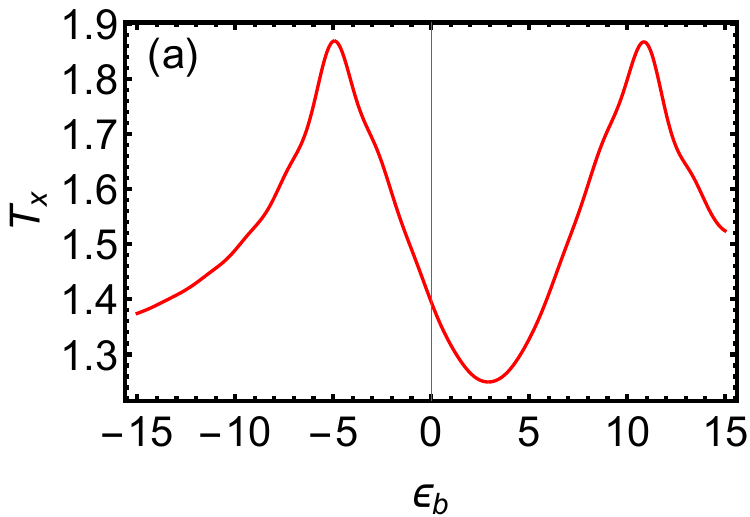}
    \phantomcaption 
    \label{Tx_a}
\end{subfigure}
\begin{subfigure}{0.9\linewidth}
    \centering
    \includegraphics[width=\linewidth]{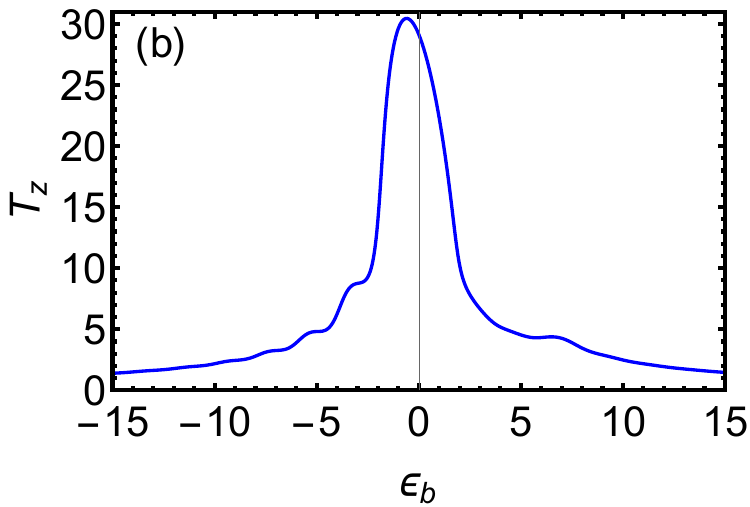}
    \phantomcaption 
    \label{Tz_b}
    \end{subfigure}
\caption[short description]{System time scale, $T_{x(z)}$ as a function of the Feshbach detuning, $\epsilon_b$. The (\red) and (\blue) curves correspond to the cases where only (a) coupling noise, $\gamma_x$ and only (b) detuning noise, $\gamma_z$ are activated, respectively.}
\label{dynamic time_detuning}
\end{figure}

If we linearize Eq. (\ref{jacobian}) about stable equilibrium point ($0,0,-1/3$) then we obtain 
\begin{equation}
(\lambda^R+\tfrac{\gamma_z}{2})
\bigg[
(\lambda^R+4\gamma_x+\tfrac{\gamma_z}{2})
(\lambda^R+4\gamma_x)
+8\tilde{g}^{\,2}
\bigg]
+k^{2}(\lambda^R+4\gamma_x)=0,
\label{detuning information}
\end{equation}
where $k=c_2-2c_1/3$. In Fig. (\ref{dynamic time_detuning}), the characteristic time $T_i$ increases as the system approaches the Feshbach resonance. The dependence on $\epsilon_b$ enters through $k$ via $c_2$. In the vicinity of resonance, the atomic and molecular states exhibit maximum hybridization, giving rise to the smallest energy gap between the two modes and the highest degree of coherence. As a result, perturbations decay most loosely near the resonance point, corresponding to the maximum relaxation time. As the spectral gap becomes minimum near about the resonance point, and since the effective gap is determined by both $\tilde{g}$ and $\epsilon_b$ on an equal footing, the coupling strength $\tilde{g}$ dominates when near resonance. As a result, $\lambda^{R}$ reaches its minimum value, leading to the maximum characteristic time scales $T_i$ for both type of noises ($\gamma_x$, and $\gamma_z$).

We now discuss the origin of the double-peak and single-peak structures observed in Figs. (\ref{Tx_a}) and (\ref{Tz_b}), respectively. Since $\gamma_x$ acts on the Feshbach-coupling channel, i.e., the atom-molecule hybridization channel, the atom-molecule modes appear as lower and upper dressed branches. In the presence of $\gamma_x$, these two modes behave differently because the atom-molecule hybridization is gradually suppressed by noise in the atom-molecule conversion channel. On either side of the resonance position, at nearly equidistant detunings, the relaxation rate is minimized, for which $T_x$ exhibits a double-peak structure. On the other hand, $\gamma_z$ induces dephasing between the atomic and molecular states. Although the coherent atom-molecule coupling remains unchanged, the coherence required to resolve the dressed modes is progressively lost. Consequently, the double-peak structure associated with the two hybridized modes gradually disappears, and a single broadened mode becomes dominant. Each dressed mode produces a distinct spectral peak, and the corresponding linewidth provides a measure of the decoherence rate. When the dephasing rate becomes comparable to the coherent coupling strength $\tilde{g}$, the spectral splitting can no longer be resolved, causing the two peaks to merge into a single broadened peak.

 Table \ref{3rd table} shows that $\epsilon_b$ is larger in magnitude than $\tilde{g}$. Therefore, in the absence of detuning noise the spectral gap is larger for the former case, resulting in comparatively smaller values of $T_x$ (Fig. \ref{Tx_a}) than those of the latter case (Fig. \ref{Tz_b}).
 
In Sec. \ref{section influence}, we identify the range of noise strengths over which noise plays a coherent role in these features by analyzing several specific time-averaged quantities.
\section{Noise-Influenced Dynamics}\label{section influence}
In Sec. \ref{intution}, we study two coherence measures as markers within the Bloch-vector formalism. In Sec. \ref{time average section}, we identify the regime in which noise promotes coherence, beyond which its conventional coherence-suppressing effect becomes dominant.
\subsection{Probing the Coherence Resonance Point through Other Physical Observables}\label{intution}
\begin{figure}[h]
    \centering
    \includegraphics[width=0.9\linewidth]{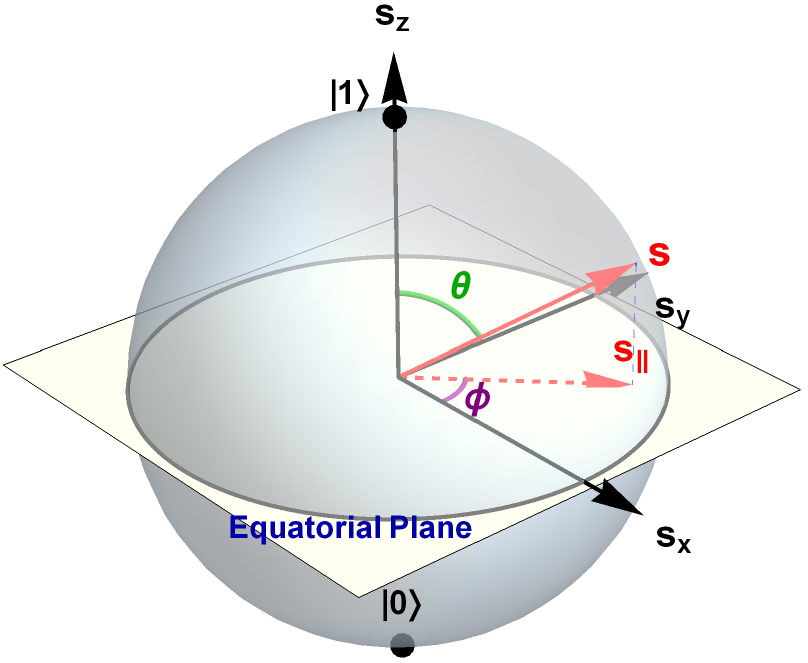}
    \caption{Bloch sphere illustrating the north ($\ket{1}$) and south ($\ket{0}$) poles, the coordinate axes $s_x$, $s_y$, and $s_z$, and the Bloch vector $\mathbf{s}$. Here, $s_{\parallel}$ denotes the projection of $\mathbf{s}$ onto the equatorial plane, while $\theta$ and $\phi$ are the polar and azimuthal angles, respectively. The equatorial plane is marked for reference.}
    \label{bloch sphere}
\end{figure}

The density matrix can be expressed as \cite{wiener,P510}
\begin{equation}
\hat{\rho} = \frac{\hat I + s_j \hat\sigma_j}{2}, \quad \text{with} \quad j \in \{x, y, z\},
\label{density matrix}
\end{equation}
where $\hat\sigma_j$ are the $(2 \times 2)$ Pauli matrices.

In its explicit matrix form, Eq. (\ref{density matrix}) becomes
\begin{equation}
\hat{\rho} = \frac{1}{2}
\begin{pmatrix}
1 + s_z & s_x - i s_y \\
s_x + i s_y & 1 - s_z
\end{pmatrix}.
\label{eigen value}
\end{equation}

\subsubsection{Purity as a Marker of the Coherence}
The purity of the state is given by \cite{P225}
\begin{equation}
P = \mathrm{Tr}(\hat{\rho}^2) = \frac{1 + \sum_j s_j^2}{2}.
\label{purity}
\end{equation} A value of $P = 1$ indicates a pure quantum state, whereas $P = 1/2$ corresponds to a completely mixed state, which is the minimum possible value of $P$, or in other words, the maximally coherent state.
\subsubsection{Husimi--Q Distribution as an Indicator of Coherence}

If $\theta$ and $\phi$ denote the polar and azimuthal angles, respectively, the Husimi-$Q$ distribution \cite{wiener,P175} is defined in terms of the angular coherent state $\ket{\theta,\phi}$ \cite{P225} as
\begin{subequations}
\begin{equation}
\begin{split}
H(\theta,\phi)=\frac{1}{\pi}\bra{\theta,\phi}\hat{\rho}\ket{\theta,\phi}
\\
=\frac{1}{2\pi}\left(1+s_z\cos\theta+s_x\sin\theta\cos\phi+s_y\sin\theta\sin\phi\right),
\end{split}
\label{husimi}
\end{equation}
\text{since, from Fig. (\ref{bloch sphere}), the components of the Bloch vector $\mathbf{s}$} {are given by $s_z = |\mathbf{s}| \cos\theta$, $s_x = |\mathbf{s}| \sin\theta \cos\phi$, and $s_y = |\mathbf{s}| \sin\theta \sin\phi$, while its projection onto the equatorial plane is $s_{\parallel} = \sqrt{s_x^2+s_y^2}$. So, }
\begin{equation}
\label{simplified}
 H(\theta,\phi)=\frac{1}{2\pi}\bigg(1+\frac{1}{\lvert \mathbf{s}\rvert}\bigg)  
\end{equation}
\text{where}
\begin{equation}
\ket{\theta,\phi}
=\cos\frac{\theta}{2}\ket{0}
+e^{i\phi}\sin\frac{\theta}{2}\ket{1}.
\label{angular}
\end{equation}
\end{subequations}
 Eq. (\ref{simplified}) shows that the Husimi-$Q$ distribution is inversely related to the length of the Bloch vector.
\subsection{Influence of Temporal Averaging on Noise Characteristics}\label{time average section}
The time-averaged value of an operator $\hat A$ is defined as \cite{mukherjee2025relaxation}
\begin{equation}
\bar A=\frac{\int_{t_i}^{t_f}A(t)\,dt}{\int_{t_i}^{t_f}dt}.
\end{equation}
We compute these averages over $0\le t\le5$, encompassing the relaxation regime. The system starts from an equal-population ($s_z^0=0$), fully coherent ($\phi=0$) state at $t=0$, whereas for $t>5$ fluctuations are largely suppressed as it approaches an atom-dominated steady state (Fig. \ref{relaxation s}). 
Though this interval does not represent an equilibrium or strictly ergodic regime, such averaging provides a useful coarse-grained measure of the system’s dominant behavior.
 
    \begin{figure}
\centering
\begin{subfigure}{0.8\linewidth}
    \centering
    \includegraphics[width=\linewidth]{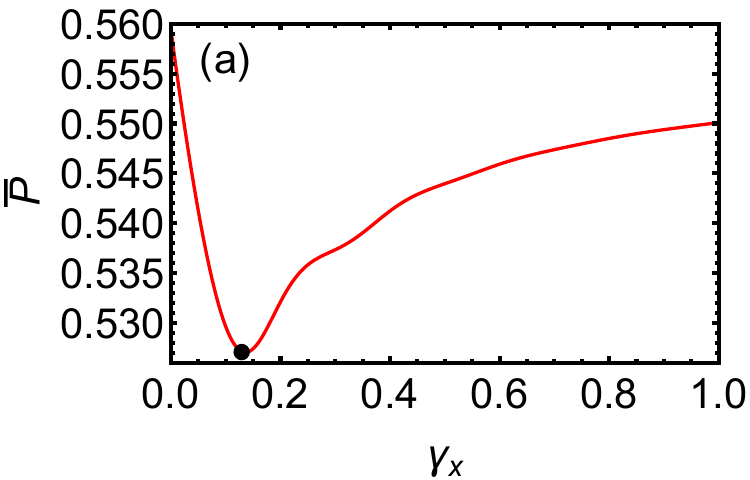}
    \phantomcaption 
    \label{purity_x}
\end{subfigure}
\begin{subfigure}{0.8\linewidth}
    \centering
    \includegraphics[width=\linewidth]{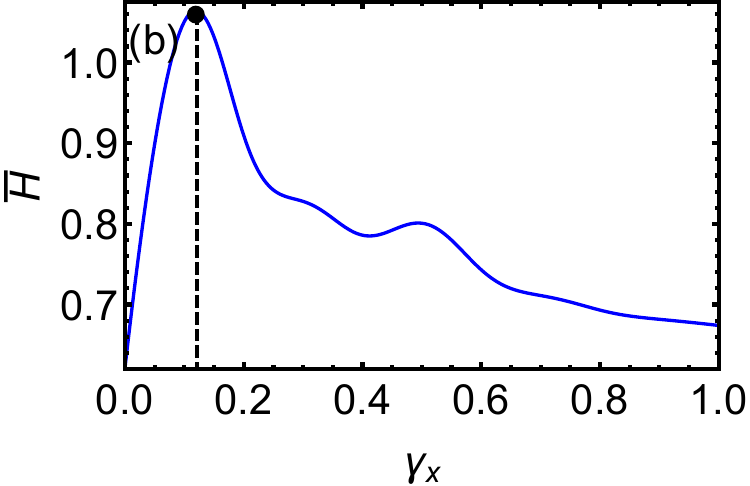}
    \phantomcaption 
    \label{husimi_x}
    \end{subfigure}
    \begin{subfigure}{0.8\linewidth}
    \centering
    \includegraphics[width=\linewidth]{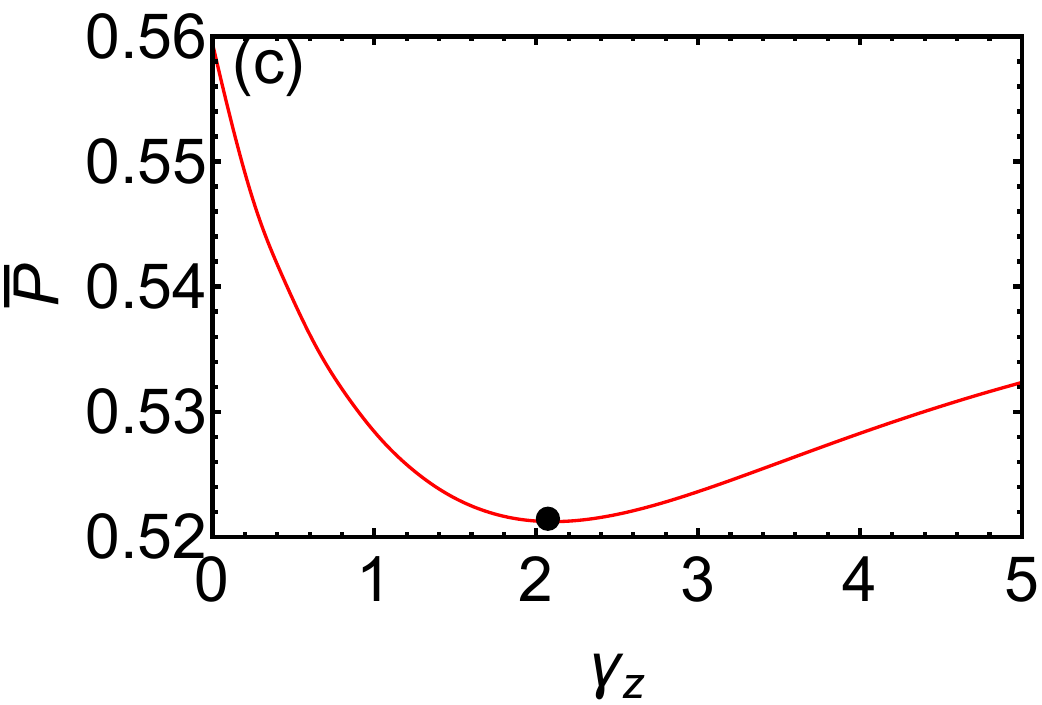}
    \phantomcaption 
    \label{purity_z}
\end{subfigure}
\begin{subfigure}{0.8\linewidth}
    \centering
    \includegraphics[width=\linewidth]{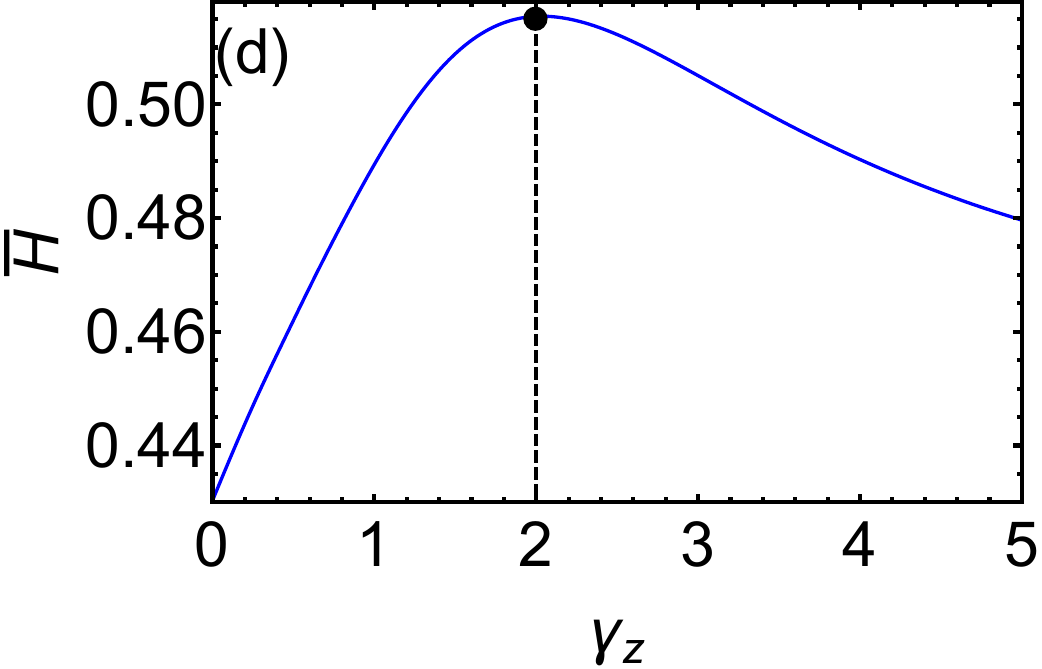}
    \phantomcaption 
    \label{husimi_z}
    \end{subfigure}
\caption[short description]{Time-averaged purity $\bar{P}$ in (a) and (c), and time-averaged Husimi-$Q$ distribution $\bar{H}$ in (b) and (d), plotted as functions of the noise strengths $\gamma_x$ and $\gamma_z$, corresponding to fluctuations in the Feshbach coupling and Feshbach detuning, respectively. Panels correspond to different values of $\gamma_x$ and $\gamma_z$, varied independently. The (\red) and (\blue) curves represent $\bar{P}$ and $\bar{H}$, respectively. The extremum value of both $\bar{P}$, and $\bar{H}$ are indicated by a ($\bullet$).}
\label{noise average}
\end{figure}

To determine the coherent resonant noise strength, $\gamma_i^{\mathrm{CR}}$, we fix the initial state and detuning and vary the noise strength. As $\gamma_i$ increases, $\bar{P}$ decreases while $\bar{H}$ increases, indicating enhanced periodicity. Beyond an optimal noise strength, however, both $\gamma_x$ and $\gamma_z$ recover their usual detrimental effects. We define the intermediate noise strength at which coherence is maximized as the coherent resonant noise strength, $\gamma_i^{\mathrm{CR}}$. For $\gamma_i>\gamma_i^{\mathrm{CR}}$, noise exerts its conventional negative effect on the system, leading to an increase in $\bar{P}$ and a decrease in $\bar{H}$.

\subsubsection{Locating the Minimum of the Time-Averaged Purity}
Fig. (\ref{purity_x}) shows that the average purity, $\bar{P}$, decreases with increasing $\gamma_x$, indicating that $\gamma_x$ plays a synchronizing role between the atomic and dimer modes. In contrast, increasing $\gamma_z$ reduces the time-averaged purity $\bar{P}$ and thus enhances the coherence of the coupled atom--molecule system [Fig. \ref{purity_z}]. Beyond a certain noise strength, however, both types of noise ($\gamma_x$ and $\gamma_z$) revert to their usual negative roles.
\subsubsection{Determination of the Maximum Time-Averaged Husimi-Q distribution}
As shown in Fig. (\ref{husimi_x}), increasing $\gamma_x$ increases $\bar{H}$ for a fixed value of $\epsilon_b$, indicating that the system becomes more coherent as the noise strength increases. In contrast, Fig. (\ref{husimi_z}) shows that increasing the detuning-noise strength $\gamma_z$ enhances fluctuations, leading to larger values of $\bar{H}$.

We now explain why a relatively larger $\gamma_z$ is required to significantly modify the dynamics compared with fluctuations introduced through $\gamma_x$. $\gamma_x$ couples directly to $\tilde{g}$ and thus directly modulates the atom-to-molecule conversion process. In contrast, $\gamma_z$ influences the energy-level separation between the atomic and molecular states. The parameter $c_2$ consists of two deterministic contributions, $U_1$ and $U_2$, in addition to the variable term $\epsilon_b$. Therefore, the effect of fluctuations in $\epsilon_b$ is mitigated by the presence of the deterministic contributions. 

In Sec. \ref{5th_section}, we investigate how the optimal noise strength depends on the initial polarization and Feshbach detuning. Moreover, we discuss the physical origin of these noise effects.
\section{Influence of Tunable Parameters and Noise Sources on the Coherence Resonance Point}
\label{5th_section}
In Sec. \ref{effect of control}, we show how the coherent noise strength can be controlled through both the initial polarization and the Feshbach detuning. In Sec. \ref{noise section}, we discuss the generation of coupling and detuning noise arising from temperature and magnetic-field fluctuations, respectively.
\subsection{Effect of Controllable Parameters on the Coherence Resonance Point}\label{effect of control}
\begin{figure}[h]
    \centering
    \includegraphics[width=0.9\linewidth]{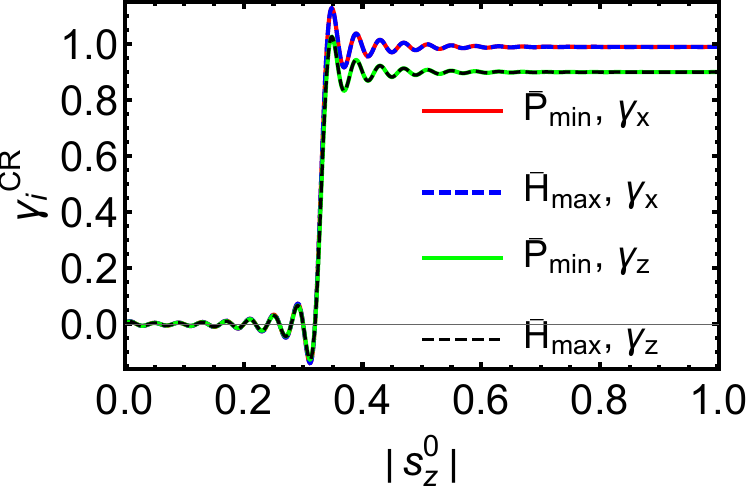}
 \caption[Short description]{Coherence resonance noise strength, $\gamma_i^{\mathrm{CR}}$, as a function of the initial polarization, $|s_z^0|$. For time-averaged Purity, $\bar{P}$, noise in the coupling strength ($\gamma_x$) and Feshbach detuning ($\gamma_z$) are represented by (\red) and (\green), respectively. For time-averaged Husimi-Q distribution, $\bar{H}$, the corresponding quantities are denoted by (\bluedashed) and (\blackdashed), respectively. The small negative region of the interpolating curve is an interpolation artifact and carries no physical significance.}
    \label{noise_polarization}
\end{figure}
\subsubsection{Impact of Initial polarization on the Coherence Resonance Point}\label{husimi section}

As $\lvert s_z^0\rvert$ increases, the Bloch vector moves farther from the equatorial plane, reducing the system's coherence. Consequently, a higher noise strength is needed to induce tunneling from the atomic to the molecular state, since the growing separation between the south and north poles of the Bloch sphere increasingly hinders coherent dynamics. As a result, the coherence-resonance point shifts to higher noise strengths. Since $s_z^{\mathrm{eq}}$ corresponds to the most stable configuration, the dynamics effectively freezes once the system reaches this point. Therefore, the largest $\gamma_i^{\mathrm{CR}}$ is required to restart the dynamics, i.e., to initiate the conversion of atoms into molecules.

On either side of this stable point, $\gamma_i^{\mathrm{CR}}$ decreases. However, on the higher-polarization side, the mixing between the two modes remains weaker; therefore, significantly larger values of $\gamma_i^{\mathrm{CR}}$ are required than on the lower-polarization side. This is shown in Fig. (\ref{noise_polarization}), where the two plots corresponding to independently activated $\gamma_x$ and $\gamma_z$ differ only slightly in the higher-polarization regime, starting from equilibrium.
\begin{figure}[h]
    \centering
\begin{subfigure}{0.9\linewidth}
    \centering
    \includegraphics[width=\linewidth]{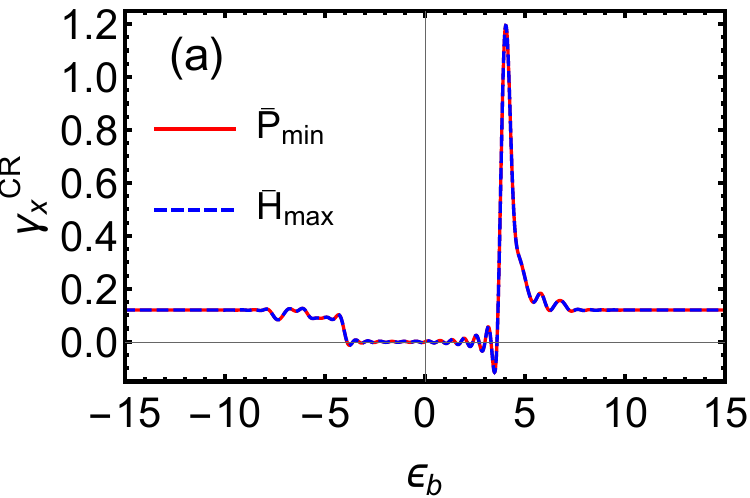}
    \phantomcaption 
    \label{gammax_b}
\end{subfigure}
\begin{subfigure}{0.9\linewidth}
    \centering
    \includegraphics[width=\linewidth]{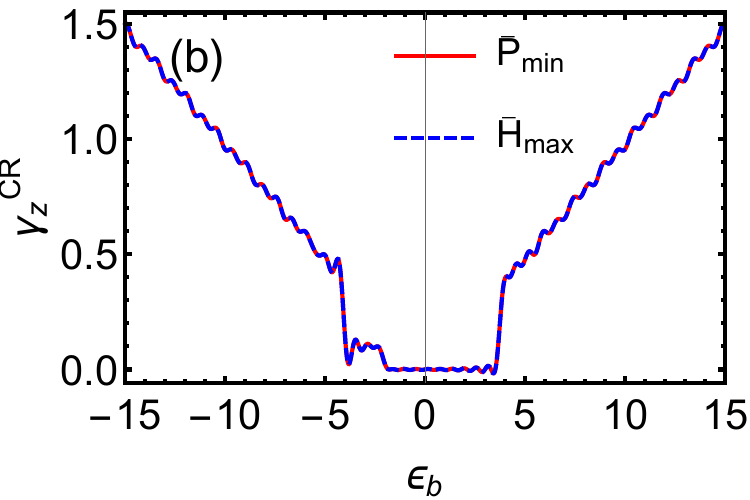}
    \phantomcaption 
    \label{gammaz_b}
    \end{subfigure}
\caption[Short description]{Coherence-resonance noise strength, $\gamma_i^{\mathrm{CR}}$, as a function of the Feshbach detuning, $\epsilon_b$. For time-averaged purity, $\bar{P}$, noise in the (a) coupling strength ($\gamma_x$) and (b) Feshbach detuning ($\gamma_z$) is represented by (\red), while the corresponding quantities obtained from time-averaged, Husimi-Q distribution $\bar{H}$ are shown by (\bluedashed), respectively. The small negative part of the interpolating curve (a) is an artifact of the interpolation and carries no physical information.}
    \label{noise_detuning}
\end{figure}

Thus, as $\lvert s_z^0\rvert$ is varied, the intrinsic time $T_i$ (Fig. \ref{dynamic time_imbalance}) and the coherent-resonance noise strength $\gamma_i^{\mathrm{CR}}$ (Fig. \ref{noise_polarization}) exhibit consistent and correlated behavior.
\subsubsection{Dependence of the Coherence-Resonance Point on Feshbach Detuning}\label{husimi section}

The dependence of the coherence-resonant noise strength on $\epsilon_b$ arises from the interplay between hybridization and fluctuations in the atomic and molecular modes. Near the Feshbach resonance ($\epsilon_b=0$), hybridization is strongest, yielding the largest effective coupling $\tilde{g}$; hence, fluctuations in $\tilde{g}$, characterized by $\gamma_x$, have the greatest impact. Since atomic--molecular conversion is maximal near resonance, only weak noise is required to induce coherence resonance. Away from resonance, the population becomes increasingly localized in either the atomic or molecular state, leading to reduced tunneling and coherence. As a result, a higher noise strength is needed to restore coherent oscillations and induce coherence resonance.

At these particular values of $\epsilon_b$, the spectral gap between the two states vanishes in Fig. (\ref{Tx_a}). In this scenario, even a minimal amount of noise, $\gamma_x$ is sufficient to overcome the barrier. Between these two minima of the spectral gap, the state separation reaches its maximum value. At this point, to overcome population freezing and achieve coherence resonance, the maximum value of $\gamma^{\mathrm{CR}}_x$ is required in Fig. (\ref{gammax_b}).

 Finite detuning diminishes the role of the effective coupling strength, leading to a increase in the resonant noise strength $\gamma_i^{\text{CR}}$ on both sides of $\epsilon_b = 0$. Under these conditions, a larger noise strength is required to reach the coherence-resonance  point, causing $\gamma_z^{\mathrm{CR}}$ to increase, as shown in Fig. (\ref{gammaz_b}). 

 Thus, the dependence of the intrinsic time scale, $T_i$ (Fig. \ref{dynamic time_detuning}), and the coherent resonance noise strength, $\gamma^{\text{CR}}_i$ (Fig. \ref{noise_detuning}), on $\epsilon_b$ remains consistent, although the two analyses are based on different physical origins.

Note that, the nonlinear nature of the coupled Bloch equations gives rise to several non-monotonic up-and-down variations in Figs. (\ref{noise_polarization}) and (\ref{noise_detuning}).
 In addition, for these two figures, $\gamma_x$ and $\gamma_z$ are required to be positive. The negative values between consecutive data points in Figs. (\ref{noise_polarization}) and (\ref{gammax_b}) result from spline overshooting and are interpolation artifacts with no physical significance.

\subsection{Origin of Noises}\label{noise section}
The condensed-thermal collision rate is taken as \cite{liu2010shapiro}
\begin{equation}
  \Gamma_x = 8\pi a_{\mathrm{eff}}^2 n_{\mathrm{th}} v_{\mathrm{th}}.  
\end{equation} This yields a diffusion rate
$\tilde{\Gamma}_x = 8\pi^3 \Gamma_x$ and consequently, $\gamma_x = \hbar \tilde{\Gamma}_x$
(Table \ref{4th table}). Here, $n_{\mathrm{th}}$ and $v_{\mathrm{th}}=\sqrt{2k_b\tilde{T}/m_a}$ denote the thermal particle density and thermal velocity, respectively where, $\tilde{T}$ is the temperature of the cloud. The effective scattering length is
$a_{\mathrm{eff}} = a_{\mathrm{bg}}
\bigg(1-\Delta B/(B-B_0)\bigg)$,
with $\Delta B$, $B_0$, $B$, and $a_{\mathrm{bg}}$ denoting the Feshbach resonance width, resonance position, applied magnetic field, and background scattering length, respectively. The corresponding identical-boson cross section is $8\pi a_{\mathrm{eff}}^{2}$ \cite{liu2010shapiro}. 
\begin{figure}
\centering
\begin{subfigure}{0.9\linewidth}
    \centering
    \includegraphics[width=\linewidth]{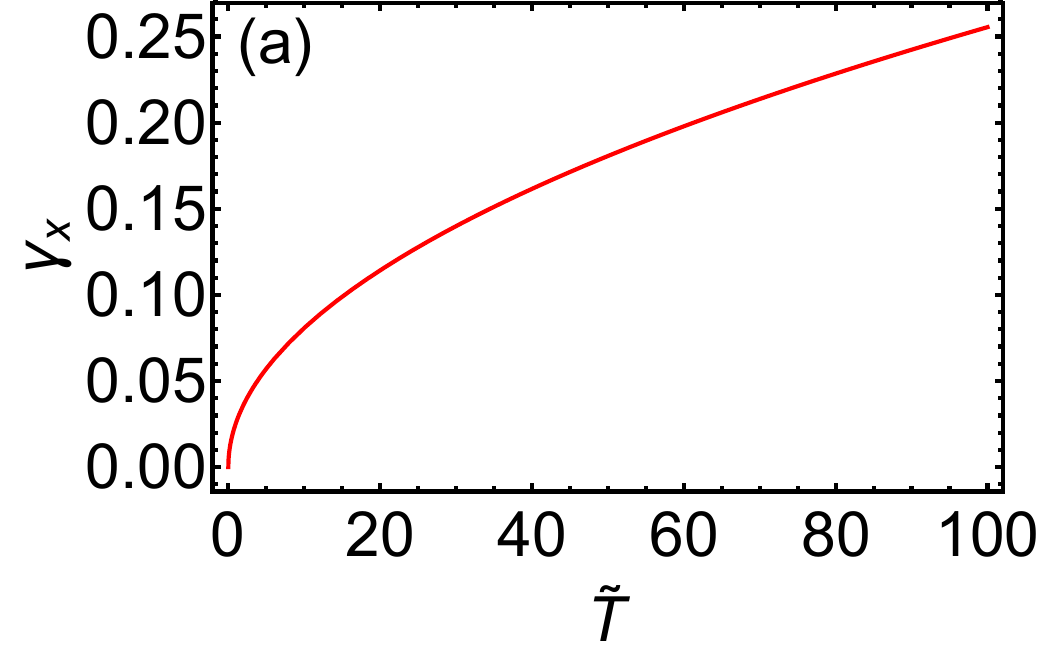}
    \phantomcaption 
    \label{temp_noise}
\end{subfigure}
\begin{subfigure}{0.9\linewidth}
    \centering
    \includegraphics[width=\linewidth]{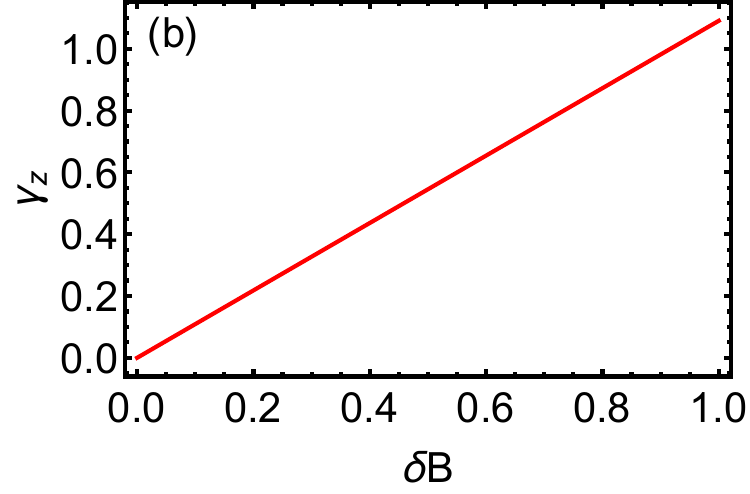}
    \phantomcaption 
    \label{magnetic_noise}
    \end{subfigure}
    \caption[short description]{(a) Dependence of the Feshbach-coupling noise strength, $\gamma_x$, on temperature, $\tilde{T}$; (b) dependence of the Feshbach-detuning noise strength, $\gamma_z$, on the magnetic-field fluctuation, $\delta B$.}
\label{origin of noises}
\end{figure}
As shown in Fig. (\ref{temp_noise}), temperatures in the $\mu$K regime are sufficient to generate the noise strength $\gamma_x$ employed in Figs. (\ref{purity_x}) and (\ref{husimi_x}).

Assuming a $10\%$ magnetic-field noise level, we set
$\delta B = 0.1\lvert B-B_0\rvert$, where $\mu_{\mathrm{co}}$ is measured in units of $\mu_B$ in Table \ref{1st table}. The resulting dephasing rate is $\gamma_z=\mu_{\mathrm{co}}\delta B$. Fig. (\ref{magnetic_noise}) shows that such magnetic-field fluctuations are sufficient to produce the values of $\gamma_z$ used in Figs. (\ref{purity_z}) and (\ref{husimi_z}). 
\section{Conclusion}\label{conclusion}
 
In this article, we investigated a system in which two bosonic atoms can bind into a bosonic molecule when coupled through a Feshbach resonance. The energy offset between the free atomic pair and the bound molecular state is defined as the detuning. In our formulation, both the detuning and the Feshbach coupling are influenced by Gaussian white noise. This two level atom-molecule system is naturally represented on the Bloch sphere: the $z$ component of the Bloch vector ($\hat{L}_z$) quantifies the population imbalance between the atomic and molecular BECs, while $\hat{L}_x$ and $\hat{L}_y$ correspond to the real and imaginary parts of the coherence, respectively. To describe the dynamics of the system, we employ the MF approach.

When the intrinsic and noise timescales become comparable, the noise can enhance the dynamics coherently. This behavior is manifested through extrema in the time-averaged purity and the Husimi-$Q$ distribution as the noise strength is varied. Consistently, the coherent noise strength decreases as the polarization is tuned away from equilibrium point of the bloch sphere and exhibits a minimum near the Feshbach resonance, where the intrinsic timescale is maximal. These observations indicate that positive noise effects emerge from the interplay between the system's intrinsic dynamical timescale and the external noise timescale.

These findings motivate further theoretical and experimental investigations of relaxation dynamics in ultracold quantum gases. Experimentally, resonantly coupled atomic-molecular BECs have been realized \cite{papp2006observation,zhang2021transition}. The coherence dynamics of coupled degenerate one-dimensional Bose gases have also been studied in detail \cite{hofferberth2007non}, while the interplay between the average fringe contrast and thermal fluctuations in one-dimensional BECs has been shown to depend sensitively on the system size \cite{hofferberth2008probing}. Relaxation phenomena have also been observed in a variety of ultracold systems, including Josephson dynamics in double-well $^{87}\mathrm{Rb}$ condensates \cite{BJJ1,zhang2021transition}, and atom-dimer relaxation processes \cite{smirne2007collisional}.

From a theoretical perspective, Coherence Resonance has been identified in the relaxation dynamics of two-state systems described by the Bloch equations and has been experimentally demonstrated in water samples \cite{viola2000stochastic}. In our context, the stronger nonlinearity of the atom-dimer Hamiltonian, compared with that of conventional double-well systems, is expected to give rise to richer dynamical behavior. The present work therefore, provides a useful framework for identifying the relevant parameter regimes and characteristic timescales that can guide future experimental studies of relaxation and stochastic-resonance phenomena in bosonic atom-molecule condensates.
\section{Acknowledgements}
The author is grateful to Raka Dasgupta for carefully reading the manuscript and for valuable discussions and comments. The author also acknowledges the University Grants Commission (UGC), Government of India, for financial support (Student ID: 201610064840). 
\setcounter{equation}{0}
\setcounter{section}{0}
\setcounter{subsection}{0}
\renewcommand{\theequation}{S\arabic{equation}}
\setcounter{enumiv}{0}

\appendix
\section*{Appendix: Initial Conditions and Parameter Estimation}\label{appendix}

\begin{table*}[t]
\centering
\captionsetup{justification=centering}
\caption{Microscopic Parameters of the System \cite{BEC9,kohler2006production,volz2003characterization}}
\begin{tabular}{|c|c|c|c|c|c|c|c|c|c|c|c|c|c|c|c|}
\hline
$m_{\text{a}}$ (kg) & $\omega$ (Hz) & $L_0$ (m) & $a_{\text{bg}}$ ($a_0$) & $a_{aa}$ ($a_0$) & $a_{bb}$ ($a_0$) & $a_{ab}$ ($a_0$) & $\mu_{\text{co}}$ $(\mu_B)$ & $B-B_0$ (G) & $B_0$ (G) & $\Delta B$ (G)&  $N_a$ (N) & $N_b$ (N)& $\tilde{T}$ $(\mu K)$ \\
\hline
$1.4\times10^{-25}$ & 100 & $10^{-6}$ & $100$ & $58$ & $10^3$ & $-180 \pm 150$ & 2 & 0.5 & 1007.4& 0.21 &2/3 & 1/3 & 10 \\
\hline
\end{tabular}
\label{1st table}
\end{table*}

\begin{table*}[t]
\centering
\captionsetup{justification=centering}
\caption{Bare Physical Inputs \cite{BEC9,li2010nonlinear}}
\begin{tabular}{|c|c|c|c|c|c|}
\hline
Quantity & $u_1\,(\text{J}\,\text{m}^3)$ & 
$u_2\,(\text{J}\,\text{m}^3)$ & 
$u_3\,(\text{J}\,\text{m}^3)$ & 
$g\,(\text{J}\,\text{m}^{3/2})$ & 
$\epsilon_b\,(\text{J})$  \\
\hline
Expression & $4\pi\hbar^2 a_{\text{aa}}/m_{\text{a}}$ & 
$4\pi\hbar^2 a_{\text{bb}}/m_{\text{b}}$ & 
$4\pi\hbar^2 a_{\text{ab}}/m_{\text{ab}}$ & 
$\sqrt{u_1 \,\Delta B \,\mu_{\text{co}}}$ & 
$\mu_{\text{co}}(B - B_0)$ \\
\hline
Value & $5.1\times10^{-51}$ & $2.6\times10^{-50}$ & $-5.1\times10^{-51}$ & $2.3\times10^{-30}$ & $9.27\times10^{-28}$\\
\hline
\end{tabular}
\label{2nd table}
\end{table*}
\begin{table}[h]
\centering
\captionsetup{justification=centering}
\caption{Scaled Effective Parameters}
\begin{tabular}{|c|c|c|c|}
\hline
Quantity & Definition & Magnitude (J) & Scaling Law \\
\hline
$U_1$ & $u_1 N_a / V$ & 1.7$\times10^{-28}$ &1 \\
\hline
$U_2$ & $u_2 N_b / V$ & $7.3\times10^{-28}$ & 4.3 \\
\hline
$U_3$ & $u_3 N / V$& $1.19\times10^{-27}$ & -7 \\
\hline
$\tilde{g}$ & $g\sqrt{N_a/V}$ &  3.4$\times10^{-28}$ &2 \\
\hline
$\epsilon_b$ & $\mu_{\text{co}}(B-B_0)$ & $1.53\times10^{-27}$ & 9.1 \\
\hline
\end{tabular}
\label{3rd table}
\end{table}
\begin{table}[h]
\centering
\captionsetup{justification=centering}
\caption{Thermal Scales of the System \cite{dalvit2000decoherence,liu2010shapiro}}
\begin{tabular}{|c|c|c|c|c|c|}
\hline
$N_{\text{th}}$ & $n_{\text{th}}$ ($\mathrm{m}^{-3}$) & $v_{\text{th}}$ (m$\text{s}^{-1}$) & $\Gamma_x$ (KHz) & $\tilde{\Gamma}_x$ (MHz) & $\gamma_x$ (J) \\
\hline
$10^3$ & $5\times10^{19}$ & $1.4\times10^{-2}$ & $0.5$ & $0.1$ & $1.3\times10^{-29}$ \\
\hline
\end{tabular}
\label{4th table}
\end{table}

The scattering parameters $a_{\text{aa}}$ and $a_{\text{bb}}$ describe atom–atom and molecule–molecule interactions, respectively, whereas $a_{\text{ab}}$ represents the atom–molecule scattering length. As listed in Table \ref{1st table}, these scattering lengths are given in units of the Bohr radius $a_0$. The reduced mass of the atom–molecule system is defined as $m_{\text{ab}} = m_a m_b/(m_a + m_b)$, with $m_a$ and $m_b$ denoting the atomic and molecular masses, respectively. The characteristic length of the harmonic trap is given by \cite{BEC9} $L_0 = \sqrt{\hbar/m_a\omega}$, where $\omega$ denotes the trap frequency. The atomic, and effective molecular particle numbers $N_a=N(1-s_z)/2$ and $2N_b=N(1+s_z)/2$ can be expressed in terms of the total number $N$ and the polarization $s_z$. For $s_z=0$, both quantities depend solely on $N$. Accordingly, the trap volume can be estimated as $V \sim L_0^3 = 2 \times 10^{-17}\mathrm{m}^3$. The corresponding bare and effective interaction strengths are provided in Tables \ref{2nd table} and \ref{3rd table}, respectively. The details of the noises considered are provided in Table \ref{4th table}.
\bibliography{bibi.bib}
\end{document}